\documentstyle[multicol,aps,prl]{revtex} 

\renewcommand{\narrowtext}{\begin{multicols}{2} \global\columnwidth20.5pc}
\renewcommand{\widetext}{\end{multicols} \global\columnwidth42.5pc}
\def\al{\alpha}
\def\be{\beta}
\def\ga{\gamma}
\def\de{\delta}
\def\ep{\epsilon}

\def\et{\eta}

\def\la{\lambda}

\def\rh{\rho}

\def\ta{\tau}

\def\ph{\phi}

\def\ch{\chi}
\def\ps{\psi}

\def\De{\Delta}

\def\La{\Lambda}

\def\Ps{\Psi}

\def\cl{{\cal L}}

\def\pt#1{\phantom{#1}}
\def\prt{\partial}

\def\lsim{\mathrel{\rlap{\lower4pt\hbox{\hskip1pt$\sim$}}
    \raise1pt\hbox{$<$}}}
\def\gsim{\mathrel{\rlap{\lower4pt\hbox{\hskip1pt$\sim$}}
    \raise1pt\hbox{$>$}}}

\newcommand{\beq}{\begin{equation}}
\newcommand{\eeq}{\end{equation}}
\newcommand{\rf}[1]{(\ref{#1})}

\begin{document}

\title{Sensitivity of CPT Tests with Neutral Mesons}    
\author{V.\ Alan Kosteleck\'y}
\address{Physics Department, Indiana University, 
          Bloomington, IN 47405, U.S.A.}
\date{preprint IUHET 377, November 1997,
published in Phys.\ Rev.\ Lett.\ {\bf 80}, 1818 (1998)} 
\maketitle

\begin{abstract}
The sensitivity of experiments with neutral mesons
to possible indirect CPT violation is examined.
It is shown that experiments
conventionally regarded as equivalent
can have CPT reaches differing by orders of magnitude
within the framework of
a minimal CPT- and Lorentz-violating extension
of the standard model.
\end{abstract}

\narrowtext

Neutral-meson interferometry is a powerful tool 
for investigating the discrete symmetry CPT.
This product of charge conjugation C, 
parity reflection P,
and time reversal T
is known to be an invariance of
local relativistic quantum field theories of
point particles in flat spacetime
\cite{cpt}.
Among the various tests of CPT
\cite{pdg},
the sharpest published bounds
are obtained with the neutral-kaon system.
As an example,
the CPT figure of merit 
$r_K \equiv |m_K - m_{\overline{K}}|/m_K$
has recently been constrained to $r_K < 1.3 \times 10^{-18}$
at the 90\% confidence level 
by the experiment E773 at Fermilab
\cite{e773,fn00}.

In neutral-meson interferometry,
bounds on CPT violation 
are extracted using a phenomenological description
of the meson time evolution.
Denote by $P^0$ any of the possible neutral mesons 
$K^0$, $D^0$, $B_d^0$, $B_s^0$
produced using the strong interaction,
and combine the Schr\"odinger wave functions of $P^0$
and its opposite-flavor antiparticle $\overline{P^0}$ 
into a two-component object $\Ps$.
Then, 
the time evolution of $\Ps$ 
is governed by a 2$\times$2 effective hamiltonian $\La$
through the equation $i\prt_t \Ps = \La \Ps$. 
Off-diagonal components of $\La$ drive
flavor oscillations between $P^0$ and $\overline{P^0}$. 

Two possible kinds of CP violation can be studied 
within this formalism.
The one usually considered 
involves T violation with CPT invariance
and is controlled by a parameter $\ep_P$.
In the kaon system,
for example,
a nonzero value of $\ep_K$
is well established
\cite{pdg}.
The other involves  
CPT violation with T invariance.
It is controlled by a complex parameter 
$\de_P\approx \De \La/\De\la$,
where $\De\La \equiv (\La_{11} - \La_{22})/2$
is half the diagonal-element difference in $\La$
and $\De\la$ is the eigenvalue difference.
In the kaon system,
a bound on $r_K$ constrains $\de_K$. 

The parameter $\de_P$ can be bounded experimentally
whether or not a nonzero value has any theoretical basis.
However,
a framework for CPT violation
based on conventional quantum field theory does exist.
The idea is that apparent low-energy CPT and Lorentz breaking
might arise spontaneously
within a more fundamental theory
that is otherwise CPT and Lorentz invariant
\cite{kps}.
Any apparent breaking 
at the level of the standard model
would then merely reflect a feature of the vacuum 
rather than a fundamental property of the theory
\cite{fn0}.
Potentially observable effects 
within this general framework
have been studied 
in neutral-meson systems
\cite{kp,ck2,kvk,exptb},
in QED 
\cite{bkr},
and in baryogenesis
\cite{bckp}.

Apparent Lorentz and CPT violation
of this type can be incorporated in a general extension of 
the minimal SU(3) $\times$ SU(2) $\times$ U(1) standard model
that preserves gauge invariance and renormalizability 
\cite{ck}.
In the underlying theory, 
the spontaneous breaking 
generates constant background expectation values as usual,
but the fields involved are 
Lorentz tensors instead of Higgs scalars. 
In the standard-model extension,
nonzero expectation values appear as coupling constants
with Lorentz indices.
For example,
an expectation value $a_\mu$
would allow a CPT- and Lorentz-violating term
$- a_\mu \overline{\ps} \ga^\mu \ps$
for a fermion $\ps$.

The present work investigates the sensitivity 
of neutral-meson experiments 
to indirect CPT-violating effects produced 
in the standard-model extension
\cite{fn1}.
Most of the theoretical considerations apply 
to any of the four neutral-meson systems.
For definiteness,
in the discussion of CPT tests
some emphasis is placed on the E773 experiment mentioned above.
The results are of immediate interest for CPT tests 
because at present 
the framework of the standard-model extension
seems to be the only available 
consistent theoretical basis for a nonzero $\de_P$
within conventional quantum field theory
\cite{fn2}.

The first step is to obtain 
an explicit expression for $\de_P$
within the standard-model extension.
A key point is that the parameter $\de_P$ 
must be C violating but P and T preserving.
This is because the strong-interaction states 
$P^0$, $\overline{P^0}$
are eigenvectors of parity 
with the same eigenvalue,
so the linear combinations 
forming the physical eigenstates $P_S$, $P_L$ of $\La$
are parity eigenstates too.
Parity is therefore preserved
during the time evolution of a neutral-meson state,
so any CP violation appearing in $\La$ 
is really C violation with P invariance.

In the lagrangian $\cl$ for the standard-model extension,
the parameters controlling the Lorentz and CPT violation 
are assumed suppressed by the (small) dimensionless ratio
of the relevant light energy scale to the Planck scale
\cite{ck}.
Thus,
only contributions linear in these parameters 
could produce observable CPT violation 
in experiments with neutral mesons. 
Also,
since $\De \La$ is flavor-diagonal,
any term in $\cl$ 
with both CPT breaking and flavor changing 
would affect $\de_P$ at most as the square 
of a small parameter and hence can be disregarded.

Remarkably,
an inspection shows that only one type of term
in the standard-model extension is flavor diagonal
while violating C but preserving P and T.
For each quark field $q$
it has the form $- a^q_{0} \overline{q} \ga^0 q$,
where $a^q_{0}$ 
is the zeroth component of a background expectation value 
$a^q_{\mu}$
that varies with the flavor $q$
(cf.\ the usual Yukawa couplings).
Thus,
at first order in perturbation theory
the diagonal elements of $\La$
depend only on $a^q_{0}$.

This result is of particular interest
both because no other 
CPT- and Lorentz-violating expectation values appear 
and because to date no other experiments 
sensitive to $a^q_{\mu}$ have been identified.
Note that
higher-order corrections from conventional interactions 
do not change the result. 
Pure strong or electromagnetic corrections 
preserve C, P, T
and therefore at most can modify
the magnitude of the contributions proportional to  
$a^q_{0}$.
Any weak-interaction corrections
violating C and P while preserving net flavor 
would be suppressed by several orders of magnitude.
Also,
possible CPT violation in the gauge sector 
is expected to be small \cite{ck}
and in any case must appear as a higher-order correction here.

As a result of CPT violation,
the parameter $a^{q_1}_{0}$
for the valence quark $q_1$  
in the $P^0$ meson
contributes with opposite sign
to $a^{q_2}_{0}$
for the antiquark $\overline{q_2}$.
This means that 
$\De\La \propto \De a_0 \equiv 
(a^{q_2}_{0} - a^{q_1}_{0})$
at leading order
\cite{fn3}.
The proportionality constant 
can be found in perturbation theory 
and is approximately one 
\cite{kp},
so from the definition of $\de_P$ one finds
\beq
\de_P \approx i 
\sin\hat\ph 
\exp(i\hat\ph) 
\De a_0/\De m
\quad ,
\label{a}
\eeq
where 
$\De m \equiv m_L- m_S$ and 
$\De\ga \equiv \ga_S- \ga_L$ 
are the mass and decay-rate differences,
respectively,
between $P_L$ and $P_S$,
and where $\hat\ph\equiv \tan^{-1}(2\De m/\De\ga)$.
A subscript $P$ is understood on all these quantities.

The result \rf{a},
valid at leading order
in all Lorentz-violating parameters
in the standard-model extension,
holds for $P^0$ mesons at rest 
in the (oriented) inertial frame 
in which $\De a_0$ is specified.
In considering the effects of rotations and boosts
on this result,
one must keep distinct 
transformations of the observer (laboratory) frame 
from changes of the momentum or orientation of a particle 
within a given observer frame.
The former are conventional Lorentz transformations,
and full covariance is maintained 
because the background fields are
perceived as changed when the observer frame changes.
In contrast,
changes of the particle momentum or orientation 
leave unaffected the background values,
producing (small) apparent changes 
in the intrinsic properties of the particle.

Equation \rf{a} shows that the result  
of a CPT test with mesons at rest in the laboratory
is explicitly rotationally invariant.
However,
no experiments with mesons at rest have been performed.
An approximation is provided by the Cornell CLEO experiment,
which involves (correlated) $B$ mesons
traveling at only about 6\% of lightspeed.
A measurement of $\de_B$ in this experiment
might therefore approximately bound $\De a_0$ 
via Eq.\ \rf{a}.

Most experiments involve relativistic mesons.
This has implications for the CPT reach 
because the $\La$ formalism 
is obtained from nonrelativistic quantum mechanics,
and so for a boosted meson 
$\de_P$ is defined in the comoving frame.
To obtain the analogue of Eq.\ \rf{a} 
valid for relativistic mesons,
one can take advantage of the covariance 
of the standard-model extension under observer boosts.
Suppose a particle at rest in the laboratory frame
is described by a lagrangian including the term 
$- a^q_{\mu} \overline{q} \ga^\mu q$,
as before.
Then,
another particle with momentum related to the first
by an inverse boost $(L^{-1})^\mu_{\pt{\mu}\nu}$
is described in the same frame by  
$- a^q_{\mu} (L^{-1})^\mu_{\pt{\mu}\nu} \overline{q} \ga^\nu q$,
since the background value $a^q_\mu$ is fixed.
The lagrangian describing this boosted particle 
in a comoving observer frame is obtained
by performing an observer boost with $L^\mu_{\pt{\mu}\nu}$,
under which both the background values and the fields transform.
The result is a term
$- a^{q\prime}_{\mu} \overline{q} \ga^\mu q$,
where 
$a^{q\prime\mu}= L^\mu_{\pt{\mu}\nu} a^{q\nu}$. 
Thus,
in the comoving frame 
in which the $\La$ formalism is valid and $\de_P$ is defined,
the CPT-violating physics is controlled by 
$a^{q\prime}_0$ instead of $a^q_0$. 
For a 4-velocity $\be^\mu \equiv \ga(1,\vec\be)$,
one has $a^{q\prime}_0= \be^\mu a^{q}_{\mu}$ and hence 
\cite{fn4}
\beq
\de_P \approx i \sin\hat\ph \exp(i\hat\ph) 
\ga(\De a_0 - \vec \be \cdot \De \vec a) /\De m
\quad ,
\label{b}
\eeq
where $\De\vec a \equiv \vec a^{q_2} - \vec a^{q_1}$.

Since the expressions for $\de_P$ 
in Eqs.\ \rf{a} and \rf{b} 
have the same phase,
the real and imaginary parts of $\de_P$
are scaled in the same way
when a meson is boosted in the laboratory.
However,
Eq.\ \rf{b} shows that 
there is an overall multiplicative factor of $\ga$
acting to enhance the CPT-violating effect.
Moreover,
the observed CPT violation 
for relativistic mesons
depends not only on $\De a_0$
but also 
on the angle $\al$ between $\vec \be$ and $\De \vec a$
and on the magnitudes 
$\be$ of $\vec\be$ and $\De a$ of $\De \vec a$.
Since $|\be\cos\al|<1$,
the factor involving $\De a$ 
is always suppressed relative to $\De a_0$,
although the combination
$\De a_0 - \vec \be \cdot \De \vec a$
may be larger or smaller than $\De a_0$.
Note that if spontaneous Lorentz breaking
generates only 0-component expectation values
as seen in the laboratory frame,
then $\De \vec a = 0$ for a meson at rest 
and so the value of $\de_P$ 
is enhanced exactly by a factor $\ga$
for a boosted meson.
If instead \it no \rm
pure 0-component expectation values are generated,
then $\De a_0 = 0$ for a meson at rest,
$\de_P$ depends on $\ga \be\De a \cos\al$,
and a meson boost is \it necessary \rm to observe any effect. 
Finally,
if $\De a > \De a_0$,
then there is a hyperbola in $\be$-$\cos\al$ space
along which CPT violation is exactly cancelled 
and near which significant suppression 
could reduce the CPT reach of certain experiments.

In an ideal case,
increasing the statistics by a factor $N$
would improve the CPT reach by a factor
$\sqrt{N}$.
However,
the variation of $\de_P$ with $\ga$
provides an alternative possibility.
Suppose for simplicity that data at fixed $\al$
are obtained from an experiment 
in a regime where $\de_P\propto\ga$
holds to a good approximation.
Then,
the CPT reach of the experiment can be doubled
either by quadrupling the statistics
or by doubling the boost.

One implication is that the bounds on $\de_P$ 
reported from different experiments may 
be inequivalent.
Experiments with comparable statistical precision
can involve mesons with very different boosts
and hence may have very different CPT sensitivity. 
In some experiments,
the boost factor is large.
For example,
the $B$ mesons in the proposed LHC-B experiment at CERN
are expected to have
$\overline{\ga} \simeq 15$.
Similarly,
the E773 experiment at Fermilab
involves kaons with mean value 
$\overline{\ga} \simeq 10^2$.
The figure of merit $r_K$ bounded in this experiment 
is proportional to $\de_K$,
so in the above scenario 
the attainable sensitivity to CPT violation 
is about two orders of magnitude better than 
the limit on $r_K$ would naively suggest.

Many experiments with neutral mesons
involve a distribution of momenta.
If CPT violation were indeed detectable in such cases,
then the predicted momentum dependence of $\de_P$
would provide a striking signal.
The presence of a momentum spectrum
also has implications for 
the extraction of a CPT bound.
Consider first an experimental asymmetry $A$ 
that is directly sensitive to a linear combination
of the real and imaginary parts of $\de_P$.
An example
is the fully integrated asymmetry $A_f$
for uncorrelated neutral-$B$ mesons
described in Ref.\ \cite{kvk}.
In a regime where $\de_P\propto \ga$ holds
and $\al$ is constant,
$\de_P$ and hence $A$ 
both scale linearly with $\ga$.
The mean value $\overline{A}$ of the asymmetry
is then proportional to $\overline{\ga}$,
determined by the form 
of the normalized meson-momentum spectrum. 
For example,
$\ga \propto p$ approximately
for large momentum magnitudes $p$,
so in this case
the effective meson momentum determining 
the CPT reach 
is just the mean momentum $\overline{p}$
of the distribution
\cite{fn5}.

Most experiments constrain $\de_P$
either through fits to time-dependent asymmetries
or from the measurement of other quantities.
For example,
the E773 experiment measures 
$\De m$, the $K_S$ lifetime $\ta_S$, 
and the usual two phases 
$\ph_{+-}$, 
$\De\ph\equiv \ph_{00} - \ph_{+-}$
related to ratios of amplitudes 
for $2\pi$ decays:
$A(K_L \to \pi^+\pi^-)/A(K_S \to \pi^+\pi^-)
\equiv |\et_{+-}|\exp(i\ph_{+-})$,
and similarly for $2\pi^0$ decays.
Defining 
$\ch \equiv \ph_{+-} - \hat\ph + \De\ph/3$,
a bound on 
$|\de_K| \approx |\et_{+-}| |\ch|$
can be extracted 
using the established value
\cite{pdg}
of $|\et_{+-}|$,
along with a bound on 
$r_K \approx 2 \De m |\de_K|/m_K \sin\hat\ph$.
It can be shown that for small CPT violation
a scaling of $\de_K$ 
by $\ga_2/\ga_1$ occurring when the
meson boost is changed from $\ga_1$ to $\ga_2$
would arise largely from
a corresponding additive change to $\ph_{+-}$
by an amount $[(\ga_2/\ga_1) - 1]|\ch|$,
while the other measured
quantities remain essentially unchanged.
In this case,
CPT violation would manifest itself 
as a momentum dependence in the observed value of $\ph_{+-}$
\cite{fn6}.

Since precision constraints on $\de_P$
require high statistics,
experiments are performed over many days.
For example,
the E773 data were collected over several months
\cite{e773}.
Effects of the Earth's motion 
on the meson velocities and orientations
must therefore be considered.

The parameter $\De a_\mu$
is constant in any special-relativistic inertial frame.
In a frame in the solar neighborhood,
the velocity of a laboratory 
on the Earth's surface is nonrelativistic,
so any associated effects can be neglected.
However,
the rotation of the Earth about its axis
introduces a time variation 
in the apparent orientation 
of $\De\vec a$ in an Earth-based laboratory
\cite{fn7}.
If in an experiment the data are taken with time stamps,
then it is possible in principle  
to search directly for such variations.
The appropriate expression for $\de_P$ is Eq.\ \rf{b}
with an apparent time dependence
for the component of $\De\vec a$
lying in the Earth's equatorial plane.

Useful bounds can also be obtained
without time binning.
If,
for example,
the data are taken with a fixed orientation of $\vec \be$
in the laboratory frame
and are uniformly distributed in time over a rotation period,
then the only net surviving CPT-violating effect
arises from the component of 
$\vec \be \cdot \De \vec a$
parallel to the Earth's rotational axis.
As an example of this geometrical effect,
if the plane of the experiment is tangent 
to the Earth at latitude $\ps$
and $\ch$ is the angle made by $\vec\be$
relative to the rotational north pole in this plane,
then 
$\de_P\propto 
\ga (\De a_0 - \be \cos\ps\cos\ch \De a_\parallel )$,
where $\De a_\parallel$ is the component 
of $\De \vec a$ parallel to the Earth's rotational axis.
If by mischance $\De\vec a$ is orthogonal 
to the Earth's rotational axis,
any experiments performed over many rotation periods 
are sensitive only to 
$\de_P \propto \ga \De a_0$.
The latter also holds
for any $\De\vec a$
if the mesons are boosted solely east-west.

For the E773 experiment,
$\be\cos\ps\cos\ch \simeq 0.6$.
An estimate of the attainable CPT bound 
on the quantities $\De a_\mu$ is then 
$|\De a_0 - 0.6 \De a_\parallel | \approx m_K r_K/2\ga
\lsim 10^{-20}m_K$.
The scale of this bound is comparable to
the ratio of the kaon mass to the Planck mass.
Note that time binning could also be used 
to investigate the possibility of time variations,
as suggested above.
In any case,
an improved bound may be feasible 
in the near future using data from 
the KTeV experiment at Fermilab
or the NA48 experiment at CERN.

The E773 experiment involves collimated beams of
uncorrelated boosted mesons,
but other types of experiment also exist.
Boosted mesons traveling in different directions
can be produced in a collider.
For example,
uncorrelated $B$ mesons with velocity orientation
throughout most of the $4\pi$ range
are accessible to LEP detectors at CERN.
In experiments of this type,
CPT studies allowing for angular dependence of $\de_P$
by binning the data in angle slices
could in principle allow separate extraction
of bounds on $\De a_0$ and $\De\vec a$.
The effects may be partially reduced 
by averaging due to the Earth's rotation.
For example,
for the special case of an integrated
uniform meson distribution 
transverse to the beam,
the angular dependence of the CPT violation 
is reduced by the time-averaged cosine 
of the angle between $\De\vec a$ and the beam. 
This effect could be avoided by time binning.

Another class of experiments
involves correlated $P_S$-$P_L$ pairs
from quarkonium decays.
The unboosted case
(symmetric factory)
produces primarily a line spectrum in the momentum,
so although a (small) $\ga$ scale factor may appear
there is essentially no momentum dependence.
However,
the $4\pi$ coverage
provides potential sensitivity 
to possible angular dependence.
Note that the values of $\de_P$ for the two mesons
in each correlated pair are typically different
because they travel in opposite directions.
Also,
the effect of the Earth's rotation 
must again be allowed for in the analysis.
In contrast,
the boosted case
(asymmetric factory)
produces a meson-momentum spectrum,
so a combination of momentum and angular dependence
can govern the CPT-violating effects.

This paper has considered 
the boost and orientation dependences 
appearing in the CPT-violating parameter 
$\de_P$ within the context 
of the Lorentz-violating standard-model extension.
A related analysis for the 
usual CP-violating parameter $\ep_P$
shows that it too could acquire contributions
from Lorentz-violating
terms in the standard-model extension
that involve C and T violation but preserve P (and CPT).
However,
the effect would require second-order 
flavor-changing contributions,
and hence it is expected to be suppressed relative
to the contributions to $\de_P$ discussed above.
Moreover,
unlike the case of $\de_P$,
conventional contributions to $\ep_P$ from low-energy physics
can arise,
so any Planck-scale effect may be masked.
It therefore appears unlikely 
that observable boost or orientation dependence for $\ep_P$ 
would arise in the context of the present framework
\cite{fn8}.

This work is supported in part 
by the Department of Energy
under grant number DE-FG02-91ER40661.

\end{multicols}

\begin{references}

\bibitem{cpt}
The discrete symmetries C, P, T are discussed, 
for example, in
R.G. Sachs,
{\it The Physics of Time Reversal}
(University of Chicago Press, Chicago, 1987).

\bibitem{pdg}
See, for example,
R.M.\ Barnett 
{\it et al.},
Review of Particle Properties,
Phys.\ Rev.\ D {\bf 54} (1996) 1.

\bibitem{e773}
B. Schwingenheuer 
{\it et al.},
Phys.\ Rev.\ Lett.\ {\bf 74} (1995) 4376;
R.A.\ Briere, 
Ph.D.\ thesis,
University of Chicago, June, 1995;
B.\ Schwingenheuer,
Ph.D.\ thesis,
University of Chicago, June, 1995.

\bibitem{fn00} 
This bound holds for negligible direct CPT violation
in decay amplitudes.
See Ref.\ \cite{e773} and 
E. Shabalin, 
Phys.\ Lett.\ B {\bf 369} (1996) 335.

\bibitem{kps}
For example,
this might occur in the context of string theory.
See
V.A. Kosteleck\'y and S. Samuel,
Phys.\ Rev.\ Lett.\ {\bf 63} (1989) 224;
{\it ibid.},
{\bf 66} (1991) 1811;
Phys.\ Rev. D {\bf 39} (1989) 683;
{\it ibid.},
{\bf 40} (1989) 1886;
V.A.\ Kosteleck\'y and R. Potting,
Nucl.\ Phys.\ B {\bf 359} (1991) 545;
Phys.\ Lett.\ B {\bf 381} (1996) 89.

\bibitem{fn0}
An analogy is the behavior of an electron in a crystal,
which reflects rotational (Lorentz) breaking
although the underlying dynamics is invariant. 

\bibitem{kp}
V.A.\ Kosteleck\'y and R.\ Potting,
in D.B.\ Cline, ed.,
{\it Gamma Ray--Neutrino Cosmology and Planck Scale Physics} \rm
(World Scientific, Singapore, 1993)
(hep-th/9211116);
Phys.\ Rev.\ D {\bf 51} (1995) 3923.

\bibitem{ck2}
D.\ Colladay and V. A. Kosteleck\'y,
Phys.\ Lett.\ B {\bf 344} (1995) 259;
Phys.\ Rev.\ D {\bf 52} (1995) 6224.

\bibitem{kvk}
V.A.\ Kosteleck\'y and R.\ Van Kooten,
Phys.\ Rev. D {\bf 54} (1996) 5585.

\bibitem{exptb}
OPAL Collaboration, 
R.\ Ackerstaff
{\it et al.},
Z.\ Phys. C {\bf 76} (1997) 401;
DELPHI Collaboration,
M.\ Feindt
{\it et al.},
preprint DELPHI 97-98 CONF 80 (July 1997).

\bibitem{bkr}
R.\ Bluhm, V.A.\ Kosteleck\'y and N.\ Russell,
Phys.\ Rev.\ Lett.\ {\bf 79} (1997) 1432;
Phys.\ Rev.\ D {\bf 57} (1998) 3932.

\bibitem{bckp}
O. Bertolami
{\it et al.},
Phys.\ Lett.\ B {\bf 395} (1997) 178.

\bibitem{ck}
D.\ Colladay and V.A.\ Kosteleck\'y,
Phys.\ Rev.\ D {\bf 55} (1997) 6760;
preprint IUHET 359 (1997), Phys.\ Rev.\ D, in press
(hep-ph/9809521). 

\bibitem{fn1}
Direct CPT violation in decay amplitudes is neglected here
because it is highly suppressed in the standard-model extension
and would be unobservable
\cite{kp}.

\bibitem{fn2}
For the neutral-kaon system,
it has been suggested that an unconventional quantum mechanics
in which the Schr\"odinger equation is 
replaced with a density-matrix formalism
might generate extra CPT-violating terms.
However,
the parameter $\de_K$ is unaffected.
See J.\ Ellis {\it et al.},
Phys.\ Rev.\ D {\bf 53} (1996) 3846.

\bibitem{fn3}
The appearance of the \it difference \rm $\De a_0$ 
in the physical observable is a general feature.
The parameters $a^q_{\mu}$ can be shifted by a constant
using a field redefinition,
so only their differences are observable
\cite{ck}.
Note also that the velocities and spins of the valence quarks 
are irrelevant here because their expectations vanish 
in the wave function for the $P^0$ meson at rest. 

\bibitem{fn4}
Although $\hat \ph$ and $\De m$ 
are indirectly affected by the boost,
this is at most second order in
the small CPT-violating parameters
and hence is unobservable.
Note also that the appearance of the
combination $\vec\be\cdot\De\vec a$
in Eq.\ \rf{b} is compatible with
the requirement that $\de_P$ 
be C violating but P and T preserving.

\bibitem{fn5}
Since the CPT reach improves linearly with $\ga$
but only as the square root of the statistics,
it may be possible to \it increase \rm
the experimental sensitivity
by examining only a (high-momentum) \it subset \rm
of the data available.
For the simple case of a constant distribution in $p$
in the regime where $\de_P \propto p$,
the gain is at most about 10\%.
For peaked distributions,
such as the Malensek-type distribution 
of the E773 experiment,
the loss of statistics at high momenta
typically offsets any gain from the $\ga$ factor.

\bibitem{fn6}
The actual E773 experiment
extracts $\ph_{+-}$ from an interference
pattern involving also a regeneration phase $\ph_\rh$,
which is unaffected to leading order by CPT violation.
Due to averaging effects,
some sensitivity to CPT violation might be lost
using a procedure for extracting $\ph_{+-}$ 
that disregards the momentum dependence,
although it is unlikely that the net CPT-violating
effect would vanish.
Particular care is required in determining $\De\ph$
because the experimental acceptances differ for
$2\pi^0$ and $\pi^+\pi^-$ decays,
and so differences occur in the reconstructed momentum spectra
used to extract $\ph_{00}$ and $\ph_{+-}$.

\bibitem{fn7}
The variation is approximately diurnal,
with a seasonal drift due to the orbital motion.

\bibitem{fn8}
Various authors have investigated
the possibility that 
the usual observed CP violation 
might be boost dependent:
J.S.\ Bell and J.K.\ Perring,
Phys.\ Rev.\ Lett.\ {\bf 13} (1964) 348;
S.H.\ Aronson
{\it et al.},
Phys.\ Rev.\ D {\bf 28} (1983) 495.

\end{references}
\end{document}